# A Boost Converter Design with Low Output Ripple Based on Harmonics Feedback

Haifeng Wang*, Member, IEEE

Penn State University New Kensington, PA, US

*hzw87@psu.edu

**Abstract-** Conventional boost converters are essential to connect low-voltage energy source such as battery with high voltage DC bus in Electric Vehicles due to its simple construction and high conversion efficiency. However, large output capacitor banks must be used in order to reduce the output voltage ripples when an inverter is connected to a boost converter. In a typical series hybrid electric vehicle powertrain, significant voltage and current perturbations will impact on battery system performance and result in considerable power loss. To address this issue, this paper proposes a new control strategy for ripple reduction in the dc-link of power electronic converters. In the proposed method, an observer is designed to adaptively estimate the dc-link voltage and current harmonics. The harmonic terms are multiplied by optimized gains to control the converter's duty cycle by negative feedback law. Entire control model for the proposed converter is illustrated and robustness is verified. Simulation and experiments show that original voltage and current ripple magnitude can be reduced significantly.

## 1. Introduction

Electric Vehicles such as pure EV and hybrid EV are attractive to provide environmentally friendly transportation with renewable energy. Fig .1 shows the electrical system configuration in EVs, including power DC-DC converter, inverter, motor and battery.

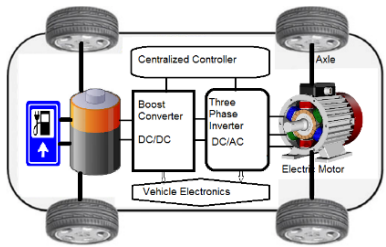

***Fig.1.*** *EV power conversion architecture*

Most of the EVs have both high power dc converter and low power converter. The first one supplies electric motor and the second one is used for ventilation, lightening and other devices [1, 2]. In electrical vehicle applications, the major load is typically an inverter followed by a motor [3, 4]. The presence of ripple and noise is an important consideration in DC converter applications [5]. The noises have two components. The first happens at the fundamental switching frequency called as ripple. The second noise occurs at the very high frequency ringing that occurs during switching transitions.

The ripple and noise levels can be sufficiently high to adversely affect other devices and parts connected to the converter if the ripple are left unfiltered. If electric vehicles are integrated with electrical grid, the charging and discharging circuits will induce extra ripples in the order of twice the grid frequency [6, 7]. Ripples in the dc-link are undesirable since they cause distortion and power loss, and generate unnecessary torque ripples. The high frequency current ripples will enter the electric vehicle's battery system to age battery and deteriorate the anions' state of equilibrium [6, 8, 9]. To reduce ripples, interleaved boost converter is widely investigated. But it needs shift several channels so that it has more components and weight as well as generates higher frequency ripples and losses due to channels' switch [10].

For typical applications where ripples are not critical, they can be filtered by adding additional ceramic capacitors to the DC-DC output end and input end. But the capacitance could be too big and vary greatly. In this case, a startup problem could arise. The output filter volume will be increased by adding the extra output capacitors to reduce the output ripples. In addition, if "LC" filtering network is applied to filter ripples, the inductance and the frequency of network should be staggered with the DC-DC frequency to avoid mutual interference. Moreover experiments indicates that the voltage ripple increases with the decreasing of the inductance in small inductance condition. Therefore, there is no doubt that simple LC filter cannot ensure the converter to meet the desired ripple requirements.

In noise sensitive applications, a need for low ripple performance is more than the converter can provide on its own. An external proportional-integral filter is recommended to be connected to the converter's output. The first step is to choose a large value of inductor. But large inductance will require a large capacitance and the response of the filter could be significantly slowed.

Adding an ultra-low-noise low-dropout regulator or LDO to the converter output is another way to get low noise and ripple in converters. But the LDO can only handle a limited current that is usually about 1A and needs required capacitors along with it. For example, a regular LDO and required components needs 46.8 mm$^2$ PCB board space and the cost is more than $5 for each LDO.

This paper proposes a method to reduce DC-LINK ripple for a typical two-stage converter inverter drive circuit that consists of a boost converter cascaded by a three-phase full bridge inverter and a motor, as depicted in Fig. 2. Compared with the previous work [11], the newly proposed 7th order autonomous LTI control system is applied to a battery powered motor drive system. This paper applies this observer based method to reduce inductor current harmonics besides DC-LINK voltage ripples. Moreover, the voltage and current harmonics frequency in the observer can be adaptively updated according to the real-time speed of the motor. This paper also compares the reduced ripple magnitude with circuit analysis to prove the effectiveness of the method.

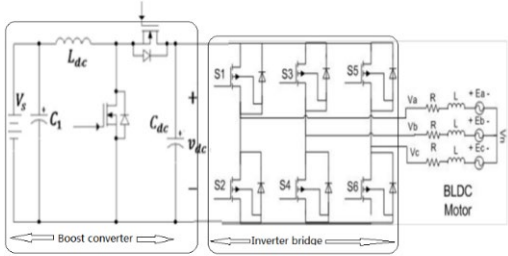

***Fig.2.*** *Circuit topology for a BLDC motor driver powered by a boost converter.*

## 2. Feedback Control design for robust reference tracking and ripple reduction

### 2.1 Extracting ripple harmonics via an observer

Suppose the above boost converter-inverter circuit is a linear-time-invariant (LTI) system represented by its state. If the system is observable, its states such as voltage and current can be reconstructed or evaluated by an observer. In what follows, the section explains how the dc-link harmonics are related to the state of the LTI system. Then the construction of observer follows from standard procedure[11].

At steady state, the dc-link voltage $v_{dc}$ can be expressed as

$$v_{dc}(t) = v_a + \sum_{n=1}^{\infty} b_n \cos(n\beta t + \phi_n) \quad (11)$$

Where $v_a$ is the average of $v_{dc}$, $b_n$ and $\phi_n$ are the magnitude and phase for the n-th order harmonic. β (measured in radians) is the frequency of the dc-link ripple and is adaptively adjusted based on motor speed ratio. The first three harmonics will be sufficient and they can be used to significantly reduce the ripple. So the paper considers only the first three harmonics. Then

$$v_{dc}(t) = v_a + b_1 \cos(\beta t + \phi_1) + b_2 \cos(2\beta t + \phi_2) + b_3 \cos(3\beta t + \phi_3) \quad (12)$$

DC-LINK voltage $v_{dc}$ and inductor current are the output of a 7th order autonomous LTI system. Define

$$S = \begin{bmatrix} 0 & 0 & 0 & 0 & 0 & 0 & 0 \\ 0 & 0 & -\beta & 0 & 0 & 0 & 0 \\ 0 & \beta & 0 & 0 & 0 & 0 & 0 \\ 0 & 0 & 0 & 0 & -2\beta & 0 & 0 \\ 0 & 0 & 0 & 2\beta & 0 & 0 & 0 \\ 0 & 0 & 0 & 0 & 0 & 0 & -3\beta \\ 0 & 0 & 0 & 0 & 0 & 3\beta & 0 \end{bmatrix} \quad (13)$$

$$G = [1 \ 1 \ 0 \ 1 \ 0 \ 1 \ 0] \quad (14)$$

And denote the state of the LTI system as $x \in R^7$, then let's analyze the voltage $v_{dc}$ first:

$$v_{dc} = Gx, \dot{x} = Sx, x(0) = x_0 \quad (15)$$

The two descriptions (1) and (2) are related via the following equations:

$$v_a = x_1(t), x_{01} = v_a \quad (16)$$

$$b_1 \cos(\beta t + \phi_1) = x_2(t), \begin{bmatrix} x_{02} \\ x_{03} \end{bmatrix} = \begin{bmatrix} b_1 \cos \phi_1 \\ b_1 \sin \phi_1 \end{bmatrix} \quad (17)$$

$$b_2 \cos(2\beta t + \phi_2) = x_4(t), \begin{bmatrix} x_{04} \\ x_{05} \end{bmatrix} = \begin{bmatrix} b_2 \cos \phi_2 \\ b_2 \sin \phi_2 \end{bmatrix} \quad (18)$$

$$b_3 \cos(3\beta t + \phi_3) = x_6(t), \begin{bmatrix} x_{06} \\ x_{07} \end{bmatrix} = \begin{bmatrix} b_3 \cos \phi_3 \\ b_3 \sin \phi_3 \end{bmatrix} \quad (19)$$

This means that each harmonic can be reconstructed by two of the state variables. Since the pair (S,G) is observable, all its states can be reconstructed from its output $v_{dc}$, which can be easily measured by AD converters. It follows that, the average of $v_{dc}$, all its harmonics and their derivatives can be extracted from $v_{dc}$ via an observer in table 1.

Based on the technology dealing with unwanted periodical signals [2, 12], the feedback control will be constructed using not only the harmonics which are represented by $x_2$, $x_4$ and $x_6$, but also their derivatives, $x_3$, $x_5$ and $x_7$.

**Table 1** Fundamental and harmonics for the DC voltage.

| | |
|---|---|
| Fundamental Component | $v_a$ |
| First Order Harmonic | $b_1 \cos \phi_1$ |
| First Order Harmonic Derivative | $b_1 \sin \phi_1$ |
| Second Order Harmonic | $b_2 \cos \phi_2$ |
| Second Order Harmonic Derivative | $b_2 \sin \phi_2$ |
| Third Order Harmonic | $b_3 \cos \phi_3$ |

| Third Order Harmonic Derivative | $b_3 \sin \emptyset_3$ |
|---|---|

### 2.2 Estimation of the harmonics via observer

Let the state of the observer be z and the input be $v_{dc}$:

$$\dot{z} = (S - LG)z + Lv_{dc} \quad (20)$$

Let the observer error be $e = z - x$, and $\dot{e} = (S - LG)e$. This implies that the observer error will go to 0 as t tends to infinity if $S - LG$ is stable. L is the observer gain. Due to measurement errors, noise, high-frequency switching ripple, the gain L should be kept within reasonable range for acceptable convergence rate and avoid over amplifying noises.

When the estimation error e is sufficiently small, the observer state z is be very close to state x of the LTI system. The harmonics of $v_{dc}$ and their derivatives can be extracted from the state z. Then state z can be used for duty cycle feedback control.

### 2.3 Using the estimated harmonics for feedback

Suppose that the current $i_L$ of the inductor $L_{dc}$ is represented by a voltage across a resistor. Let the desired dc-link voltage be $v_{ref}$. A simple feedback to achieve robust tracking is

$$D = D_0 + k_1(i_L - i_{L0}) + k_2(v_{dc} - v_{ref}) + k_3 \int (v_{dc} - v_{ref}) dt \quad (21)$$

Where D is the duty cycle for the low side MOSFET of the half-bridge for the dc-dc converter and $D_0$ is a nominal duty cycle. $i_{L0}$ is the nominal inductor current. To reduce dc-link voltage ripple, an extra term Kz is added in the above feedback control:

$$D = D_0 + k_1(i_L - i_{L0}) + k_2(v_{dc} - v_{ref}) + k_3 \int (v_{dc} - v_{ref}) dt + Kz \quad (22)$$

### 2.4 The discretized observer and digital implementation

The LTI system needs to be discretized for implementation in a digital processor. Let the sampling time be T and the discretized state is $x[k] = x(kT)$.

The discretized LTI system is

$$v_{dc}[k] = Gx[k], x[k+1] = S_d x[k], x[0] = x_0 \quad (23)$$

Where $S_d = e^{S*\frac{\omega}{1000}*t} = $

$$\begin{bmatrix} 1 & 0 & 0 & 0 & 0 & 0 & 0 \\ 0 & \cos\beta T & \sin\beta T & 0 & 0 & 0 & 0 \\ 0 & -\sin\beta T & \cos\beta T & 0 & 0 & 0 & 0 \\ 0 & 0 & 0 & \cos 2\beta T & \sin 2\beta T & 0 & 0 \\ 0 & 0 & 0 & -\sin 2\beta T & \cos 2\beta T & 0 & 0 \\ 0 & 0 & 0 & 0 & 0 & \cos 3\beta T & \sin 3\beta T \\ 0 & 0 & 0 & 0 & 0 & -\sin 3\beta T & \cos 3\beta T \end{bmatrix} \quad (24)$$

Similarly, the state observer is constructed as

$$z[k] = S_d z[k] + L_d (v_{dc}[k] - Gz[k]) \quad (25)$$

Where $L_d$ is chosen so the eigenvalues of $S_d - L_d G$ are within the unit disk to keep stable. A simple rule is to choose the desired eigenvalues within the unit disk but very close to those of $S_d$, which are all on the unit disk. Let the eigenvalues of $S_d$ be $\lambda_i, i = 1,2,..,7$. the eigenvalues of $S_d - L_d G$ at $\rho\lambda_i, i = 1,2,..,7$. with $0 < \rho < 1$. If the sampling time is very small, then $\rho$ should be very close to 1, for example, greater than 0.99.

Notice that $S_d$ is 7 by 7 matrix. This may slow down the computation in a digital signal process. To reduce computation time, the system is decomposed into one first-order system and 3 second order systems.

### 2.5 System simulation and Key performance evaluation

The proposed control system block diagram is shown in Fig.3. Each block in Fig.3 is expanded in Fig. 4 for details.

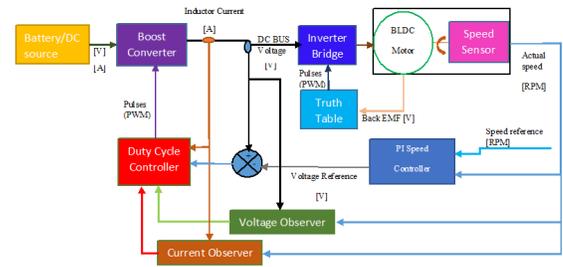

*Fig.3. Block diagram for the entire control system*

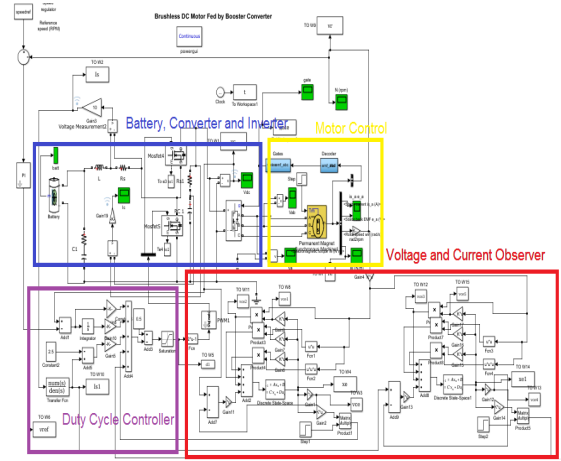

*Fig.4. Expanded block diagram of low harmonics boost converter-inverter motor control model*

The nominal operating condition was chosen at $V_s$=12V, and $V_{dc}$=28V. This requires the duty cycle for the dc-dc boost converter to be $D_0$= 0.42. And the

inductor current for $L_{dc}$ be $i_{L0}=0.9A$. The feedback control was found to be

$$D = D_0 - 0.08(i_L - i_{L0}) - 0.06(v_{dc} - v_{ref}) + \int(v_{dc} - v_{ref})dt \quad (26)$$

where $v_{ref}$ is the reference value for the dc link voltage $v_{dc}$. The coefficients were carefully designed based on a good transient performance and reasonable steady state behavior. For example, if the coefficient for $i_L-i_{L0}$ is large, the duty cycle will be noisy since the inductor current has large high-frequency switching ripple. If this coefficient is reduced, there will be larger overshoots in inductor current and $v_{dc}$. The coefficient for $v_{dc} - v_{ref}$ should be kept small due to the constraint on the duty cycle and the noises. Increasing the gain for the integrator may increase the overshoots and the ripple size. And the duty cycle D is constrained within [0, 0.8] via a saturation function as D must less than 1.

For digital implementation, the paper first constructed a discretized observer to extract the first order, the second-order harmonics and third-order as well as their derivatives from the state of a 7th order LTI system. Since the real ripple frequency is 400Hz, $\beta = 2\pi \times 400 = 2512 rad/s$.

With sampling frequency of 18 kHz, the sampling period is $T = 5.56 \times 10^{-5} s$. And the $S_d$ is expressed as below:

$$S_d = \begin{bmatrix} 1 & 0 & 0 & 0 & 0 & 0 & 0 \\ 0 & 0.9903 & -0.1392 & 0 & 0 & 0 & 0 \\ 0 & 0.1392 & 0.9903 & 0 & 0 & 0 & 0 \\ 0 & 0 & 0 & 0.9610 & -0.2756 & 0 & 0 \\ 0 & 0 & 0 & 0.2756 & 0.9610 & 0 & 0 \\ 0 & 0 & 0 & 0 & 0 & 0.9123 & -0.4066 \\ 0 & 0 & 0 & 0 & 0 & 0.4066 & 0.9123 \end{bmatrix} \quad (27)$$

The desired eigenvalues for $S_d - L_d G$ are obtained by scaling the eigenvalues of $S_d$ with 0.99[11]. Using pole placement algorithm, $L_d$ is computed as

$$L_d = \begin{bmatrix} 0.0098 \\ 0.0195 \\ 0.0019 \\ 0.0192 \\ 0.0037 \\ 0.0189 \\ 0.0047 \end{bmatrix} \quad (28)$$

As the Fig.5 shows, the dc-link voltage is decomposed into a dc component and harmonics by the built observer. Recall that the first-order and second order harmonics correspond to $z_2$, $z_4$ and $z_6$ of the observer. $z_3$, $z_5$ and $z_7$ are similar to $z_2$, $z_4$ and $z_6$ but with some phase shift.

Let's compare the original $v_{dc}$ signal and reconstructed signals from observer states. Fig. 6 (a) shows the original $v_{dc}$ (top, blue curve) and the reconstructed $v_{dc}$ (middle, red curve). The estimation error is also plotted as the black curve at the bottom of the figure. The comparison shows the effectiveness of the observer and that the 7th-order model is sufficient for the dc-link voltage.

A close up view of the comparison is shown in Fig. 6(b). It is obvious that the curve for the original $v_{dc}$ is fuzzy and thick, due to the high frequency switching ripple. In contrast, the reconstructed $v_{dc}$ is clean and ripple free, indicating that the high frequency switching ripple has been filtered out due to effect of the observer.

Since the fourth-order harmonic is negligible at steady state, the system only need feedback the state $z_2$, $z_3$, $z_4$, $z_5$, $z_6$ and $z_7$ to calculate the duty cycles. A state feedback is designed by adding two more terms to as:

$$D = D_0 - 0.08(i_L - i_{L0}) - 0.06(v_{dc} - v_{ref}) + \int(v_{dc} - v_{ref})dt - 0.3z_2 + 0.2z_3 - 0.1z_4 + 0.2z_5 - 0.03z_6 + 0.14z_7 \quad (29)$$

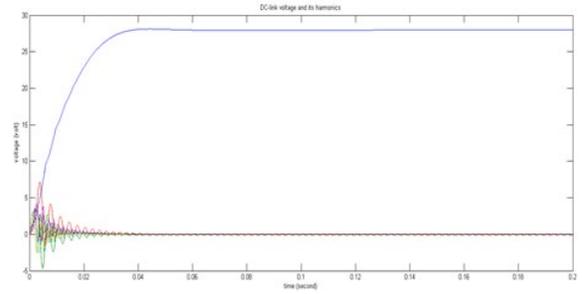

*Fig.5.* Estimated the dc-link voltage and its harmonics

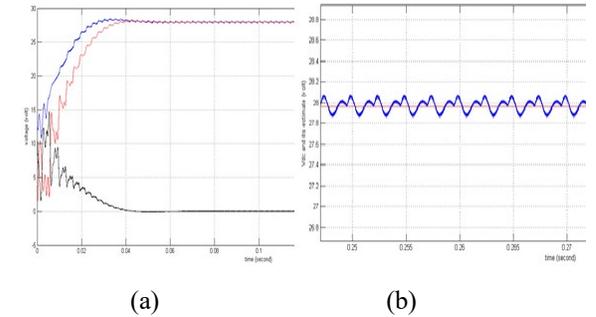

(a)            (b)

*Fig.6.* (a) reconstructed dc-link voltage, the original, and the error; (b) Close-up view of reconstructed dc-link voltage in red and the original in blue

The last two coefficients are obtained through trial and error. First, the term for $z_3$ is set to 0. By using one dimensional search and simulation, it is found that the ripple is minimum when the coefficient of $z_2$ was -0.25. Another one dimensional search with the coefficient of $z_2$ fixed at -0.25 produced a value of −0.4 for $z_3$. Then other coefficients are determined in the same way.

It should be noted that both terms for $z_2$ and $z_3$ are needed to minimize ripple size. If the term for $z_3$ is discarded, the minimal ripple size increases. This means that the derivative of the first order harmonic also helps to reduce the ripple size. The same applies to other harmonics.

In order to eliminate the output voltage overshoot and vibration at the beginning because of the observer feedback, a delay element is added to the observer feedback loop with 0.1s. To observe the effect more obviously, more load is connected to the motor. The load torque change increases the DC- link voltage and the voltage ripples. To reduce the voltage ripple, observer based control begin to work at 0.1 second.

Simulink is used to conduct the simulation of the proposed control system. The sampling rate can be chosen as 18 kHz. We use 24V DC as the output voltage of boost converter to drive 24V BLDC motor used in an experiment. The dc-link voltage generated by the boost converter controlled by two different controllers is plotted in Fig. 7. Fig.7 (a) shows the voltage ripple reduction by switching to the harmonics feedback controller. The voltage ripple peak to peak value is quickly reduced 53% from 0.37 V shown in Fig. 7 (b) to 0.17 V shown in Fig.7 (c). The response in Fig. 6 clearly shows the reduction in ripple size after the harmonic terms are fed back to the duty cycle controller after t > 0.1s.

Similar to voltage ripple reduction principle, 3-order harmonics are used in negative feedback with proper gain to reduce the input current ripple. The new observer feedback gain is gained by one dimensional search and simulation. Current ripple observer based controller begins to work at 0.2 second. The inductor current under the modified controller is plotted in Fig.8. The inductor current ripple reduction is shown in Fig.8 (a). The ripple peak to peak value is reduced 30% from 1.7 A shown in Fig. 8(b) to 1.2A shown in Fig. 8(c).

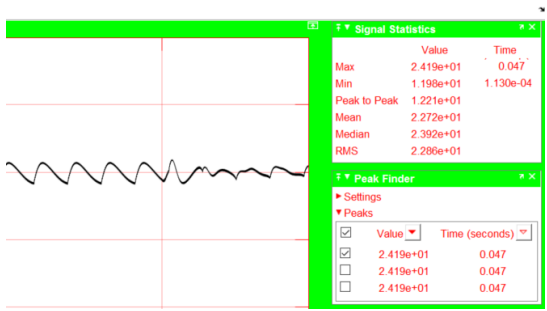

*(a) The voltage ripple reduction*

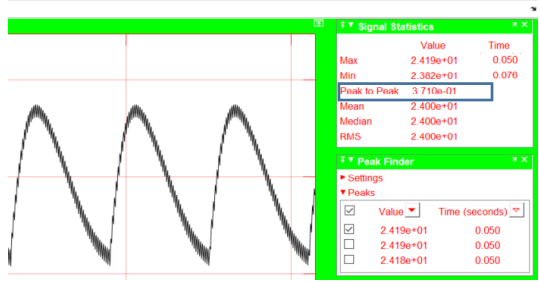

*(b) The original voltage ripple close-up view*

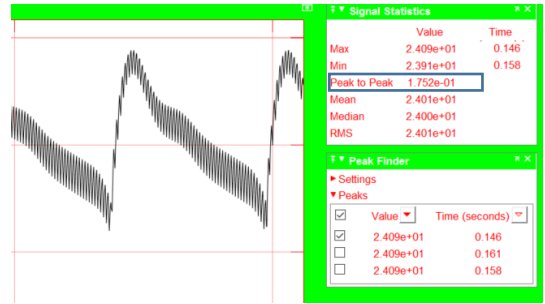

*(c) The reduced voltage ripple close-up view*

*Fig.7. Dc-link voltage ripple is reduced from 0.37 V to 0.17V.*

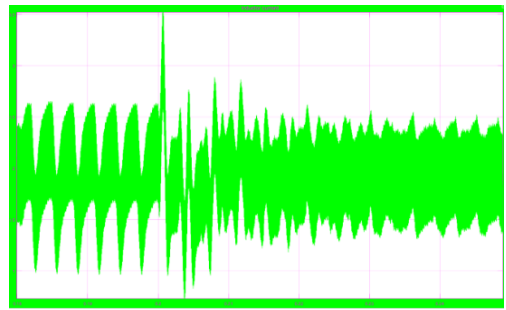

*(a) Inductor current ripple reduction*

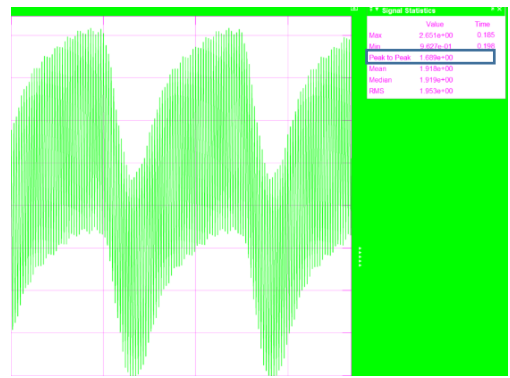

*(b) The original current ripple close-up view*

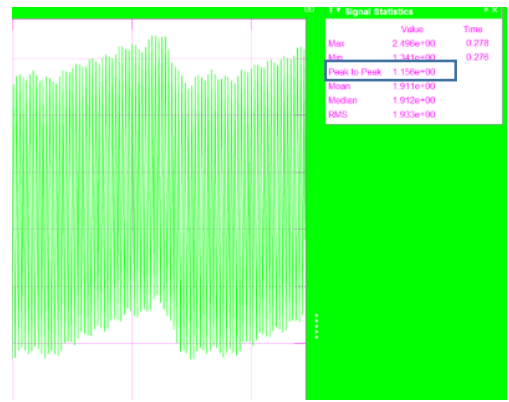

*(c) The reduced current ripple close-up view*

*Fig. 8. Inductor current ripple peak to peak value reduced from 1.688A to 1.156A*

To compare the reduced ripple magnitude in simulation with circuit analysis results, the voltage ripple and current ripple are calculated in next sections.

### 2.6 Output Voltage Ripple Analysis

To calculate a minimum output capacitance, the desired output voltage ripple percentage is set to 1% of the rated output voltage[10]. Based on Table 2, the output capacitor values for the desired output voltage ripple is as below:

$$C_{out}(\min) = \frac{I_{out(MAX)}*D}{f_S*\Delta V_{out}} = \frac{2.65*0.55}{18000*0.24} = 0.000337F = 337uF \quad (7)$$

So a 470 $uF$ capacitor is determined to be the output capacitor in the test.

The ESR of the output capacitor adds more ripples. So the total voltage ripple is given with the equation:

$$\Delta V_{out(withESR)} = ESR * \left[\frac{I_{outmax}}{1-D} + \frac{\Delta I_l}{2}\right] = 0.67V \quad (8)$$

**Table 2** Components in the simulation model

| | |
|---|---|
| $C_{out}(\min)$=337 $uF$ | minimum output capacitance |
| Iout (max)=2.65A | maximum output current |
| D = 0.55 | duty cycle |
| fs = 18 kHz | minimum switching frequency |
| $\Delta V_{out}$=0.24 V | desired output voltage ripple |
| $\Delta V_{out}$(with_ESR) = 0.67$V$ | With additional output voltage ripple due to capacitors ESR |
| ESR = 0.1ohm | equivalent series resistance of the output capacitor |
| $\Delta I_l$ =1.28A | inductor ripple current |
| $V_{in}$ = 13.9V | typical input voltage |
| Vout =24V | desired output voltage |
| L =0.00033 H | selected inductor value |

### 2.7 Input Current Ripple Analysis

The resistance of a CMOS switch in the closed position, referred to as the on-resistance or $R_{ON}$, changes depending on the input voltage. This behavior is usually undesirable and can significantly distort the input signal in some applications. A good estimation for the inductor ripple current is 20% to 40% of the output current.

$$\Delta I_{lest} = (0.2 \text{ to } 0.4) * I_{out}max * \frac{V_{out}}{V_{in}} = (0.9A \text{ to } 1.8A) \quad (9)$$

To calculate the maximum switch current is to determine the inductor ripple current.

$$\Delta I_{lmax} = \frac{V_{in}*D}{f_s*L} = \frac{13.9*0.55}{18000*0.00033} = 1.28A \quad (10)$$

In sum, the voltage ripple plotted in simulation is 0.17v (or 0.7% of output DC voltage) is better than the acceptable voltage ripple 0.67v calculated by above circuit analysis. In addition, inductor current ripple of the model is 1.16A (or 43% of output current) is also better than maximum switch current 1.28A from the above circuit analysis. In a research about the interleaved boost converter, the voltage ripple varies from 0.3% to 0.43% of output DC voltage when output power changes from 20kw to 70 kw and the inductor current ripple varies from 50% to 30% of maximum input current when output power changes from 50kw to 170 kw [13]. Their ripple to DC signal ratios, in comparison, are close to our results although the two converters' circuits and output power are different.

## 3. Experimental Verification

The experiment circuit in an electric vehicle prototype includes DSP controller, boost converter-inverter, battery and charger, motor driver and motors shown in Fig.9.

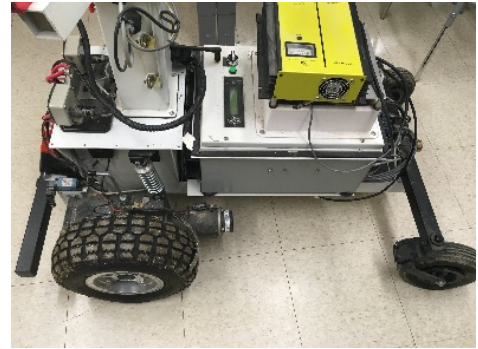

*Fig. 9.* The electric vehicle prototype in test

The detailed experiment parameters are shown in table 3 and table 4.

**Table 3** BLDC motor parameters

| | |
|---|---|
| Number of phases | 3 |
| Back EMF waveform | Trapezoidal |
| Stator phase resistance (ohm) | 0.41 |
| Stator phase inductance (H) | 0.0007 |
| Torque constant (N.M/A-peak) | 1.4 |
| Back EMF flat area (degrees) | 120 |
| Inertia (J(kg.m^2)) | 9.6e-5 |
| viscous damping (F(N.m.s)) | 1e-3 |
| pole pairs | 4 |

**Table 4.** ALM 12V7s EverSafe™ battery: lithium-ion batteries

| Electrical Characteristics at 25°C | 12V7s EverSafe™ battery |
|---|---|
| Nominal Voltage | 13.2 V |
| Nominal Capacity | 5 Ah |
| Max. Charge Voltage | 16 V |
| Recommended Float Voltage | 13.6 - 14.4 V |
| Under-voltage Limit (min) | 8 V |
| Constant Power Discharge End Voltage | 10V |
| Product Model | ALM 12V7s, Single Unit |

When motor runs at 1000 rpm, the converter output 24V voltage to power the inverter and drive the motor. Experimental data show that the boost converter output voltage ripple can be reduced by half. Fig.10 shows the voltage ripple reduction by using the harmonics feedback control. The voltage ripple peak to peak value is quickly reduced 46% from 0.67 V shown in Fig. 10 (a) to 0.36 V shown in Fig.10 (b). This reduction ratio is approximately equal to the ratio of the simulation experiment, as demonstrated in Fig. 6.

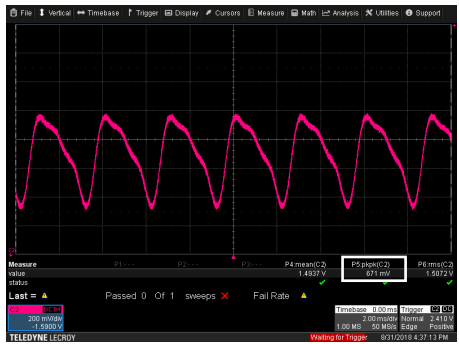

(a)

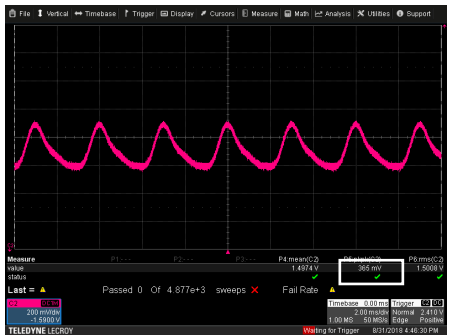

(b)

*Fig.10. Converter output voltage ripple in close-up view: (a) voltage harmonics is not fed back to the duty cycle controller; (b) voltage harmonics is fed back to the duty cycle controller*

The proposed design can also be used to reduce inductor current ripple in the boost converter. The second test uses a 0.1 ohm resistor connected in series with an inductor to measure the current flowing through the inductor. Fig.11 plots the current ripple reduction by using the harmonics feedback control. The voltage ripple peak to peak value is quickly reduced 22% from 306 mv without harmonics term feedback control in Fig.11 (a) to 239 mv with the proposed harmonics feedback control in Fig.11 (b). This ratio of ripple reduction is close to the ratio of the simulation experiment, as demonstrated in Fig. 7.

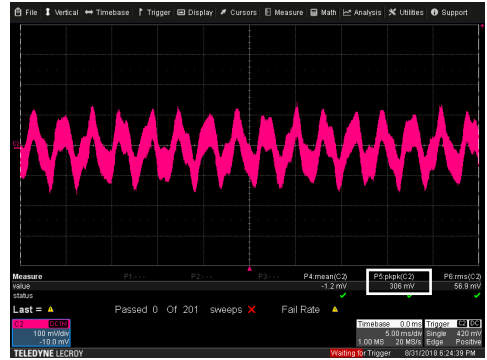

(a)

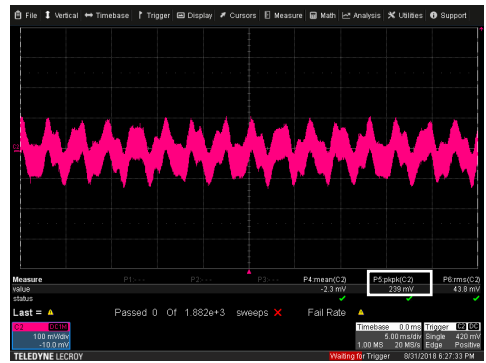

(b)

*Fig.11. Inductor current ripple sampling waveform in close-up view: The system does not use harmonics feedback in (a); the system uses harmonics feedback control in (b)*

### 4. Conclusion

The paper addressed converter ripple reduction problems for a power system driven by battery or other DC power sources. A control strategy for reduction of the dc-link ripple is proposed for a two stage boost inverter. The main idea is to use an observer to extract the harmonics of the dc-link voltage and current to use the harmonics for feedback control of the dc-dc boost converter. The motor's speed is sampled and used to adaptively adjust the frequency of the harmonics in the observer. The effectiveness of the control design methods was validated by both simulation and experiments. The results developed in this paper will find possible applications in electric, hybrid and plug-in hybrid electric vehicles, wind systems and photovoltaic systems. Since the ripple reduction is achieved by feedback control of observer states, the method can be applied to reduce other fixed frequency ripples besides voltage and current ripples.